\documentclass[12pt]{article}
\usepackage{float} 
\usepackage{hyperref}
\usepackage{booktabs}
\usepackage{algorithm}
\usepackage{algpseudocode}
\usepackage{caption}
\usepackage{newfloat} 

\usepackage{newtxtext,newtxmath}

\usepackage{graphicx}

\usepackage[letterpaper,margin=1in]{geometry}

\renewenvironment{abstract}
	{\quotation}
	{\endquotation}

\date{}

\makeatletter
\renewcommand{\fnum@figure}{\textbf{Figure \thefigure}}
\renewcommand{\fnum@table}{\textbf{Table \thetable}}
\makeatother

\usepackage{cite}

\usepackage{url}

\def\scititle{Inverse Design of Nonlinear Mechanics of Bio-inspired Materials Through Interface Engineering and Bayesian Optimization}
\title{\bfseries \boldmath \scititle}

\author{%
    Wei Zhang$^{1}$,
    Mingjian Tang$^{1}$,
    Haoxuan Mu$^{1}$,\\
    Xingzi Yang$^{2}$,
    Xiaowei Zeng$^{2\ast}$,
    Rui Tuo$^{3}$,\\
    Wei (Wayne) Chen$^{1\ast}$,
    Wei Gao$^{1,4\ast}$\and
\begin{tabular}{@{}p{\textwidth}@{}}
    \raggedright 
    \small $^{1}$J. Mike Walker $'$66 Department of Mechanical Engineering, Texas A\&M University, College Station, Texas 77843, United States\\
    \small $^{2}$Department of Mechanical Engineering, The University of Texas at San Antonio, San Antonio, Texas 78249, United States\\
    \small $^{3}$Department of Industrial \& Systems Engineering, Texas A\&M University, College Station, Texas 77843, United States\\
    \small $^{4}$Department of Materials Science \& Engineering, Texas A\&M University, College Station, Texas 77843, United States\\
    \small$^\ast$Corresponding authors. Emails: wei.gao@tamu.edu (Wei Gao), xiaowei.zeng@utsa.edu (Xiaowei Zeng), w.chen@tamu.edu (Wei (Wayne) Chen)
\end{tabular}
}

\begin{document} 

\maketitle

\newpage


\section*{Abstract}
\begin{abstract} \bfseries \boldmath
In many biological materials such as nacre and bone, the material structure consists of hard grains and soft interfaces, with the interfaces playing a significant role in the material's mechanical behavior. This type of structures has been utilized in the design of various bio-inspired composite materials. Such applications often require the materials to exhibit a specified nonlinear stress-strain relationship. A key challenge lies in identifying appropriate interface properties from an infinite search space to achieve a given target stress-strain curve. This study introduces a Bayesian optimization (BO) framework specifically tailored for the inverse design of interfaces in bio-inspired composites. As a notable advantage, this method is capable of expanding the design space, allowing the discovery of optimal solutions even when the target curve deviates significantly from the initial dataset. Furthermore, our results show that BO can identify distinct interface designs that produce similar target stress-strain responses, yet differ in their deformation and failure mechanisms. These findings highlight the potential of the proposed BO framework to address a wide range of inverse design challenges in nonlinear mechanics problems.

\end{abstract}

\newpage

\noindent
\section*{INTRODUCTION}
Many biological materials, such as nacre \cite{Barthelat2007}, bone \cite{rho1998mechanical}, limpet teeth, and fish scales \cite{browning2013mechanics, huang2019multiscale}, share a similar hierarchical structure composed of hard grains densely packed and bonded together by a thin layer of biopolymer. For instance, nacre consists of 95 wt\% aragonite and 5 wt\% organic materials. Despite the biopolymer's minimal volume fraction at the interface, it plays a significant role in the material's toughening and strengthening mechanisms by facilitating grain sliding, energy dissipation, and controlling the crack propagation \cite{Khayer2013, Barthelat2016}. Inspired by this unique material architecture, especially the brick-and-mortar structure in nacre, researchers have developed various bio-mimetic functional materials through different synthesis approaches \cite{Madhav2023, Yang2019, Peng2020}. For example, hierarchically structured fiber was fabricated by wet-spinning assembly technology, using graphene as tablets and hyperbranched polyglycerol (HPG) as adhesive, which achieved excellent strength and corrosion-resistance \cite{Hu2012, Hu2016}. Nacre-inspired multi-layered films were fabricated using an immersive layer-by-layer assembly technique for fire-retardant coating \cite{Yan2016, Pan2022} and the separation of metallic ions in aqueous solution \cite{Qin2016}.

Over the years, extensive research has aimed to model and optimize the structure of bio-inspired materials to maximize or balance strength and toughness, mimicking natural designs. In these efforts, the design parameters generally include the geometry and material properties of both the tablets and the interface. For instance, Ni et al. developed a nonlinear shear-lag model for brick-and-mortar structures, achieving simultaneous optimization of composite toughness and strength by adjusting parameters related to tablet dimensions and properties, as well as the interface parameters \cite{Ni2015}. Subsequent numerical studies revealed that the optimal trade-off between toughness and strength could be determined by tuning the aspect ratio and volume fraction of the tablets \cite{Zhou2021, Zhou2023}. Radi et al. investigated the influence of interface reinforcement on both strength and toughness \cite{Radi2020}. Al-Maskari et al. developed analytic models based on a representative volume element approach to examine the effects of tablet geometry, as well as the interface's shear stiffness and strength \cite{Al-Maskari2021}. More recently, Park et al. applied multi-objective Bayesian optimization to identify the optimal balance among strength, toughness, and specific volume by varying tablet geometry and the elastic modulus ratio of the hard and soft constituents \cite{Park2023}. 

Despite years of research focusing mainly on enhancing the strength and toughness of bio-inspired materials, many applications now require such materials to exhibit a predefined nonlinear stress-strain response. For example, nacre-inspired composite materials have recently been applied in stretchable electronics \cite{Xu2021, wang2022flexible}, where designing the material to achieve target stress-strain curves can be crucial. A key unresolved question is whether, given a predefined nonlinear stress-strain curve, one can identify the corresponding interface properties — an inverse problem that has yet to be addressed. In this study, we propose a Bayesian optimization (BO) framework to tackle this inverse design problem. BO, a sequential model-based approach \cite{Shahriari2016}, has been increasingly utilized to solve inverse problems in material design \cite{Kuszczak2023, Xue2020, Park2023}. At its core, the BO framework consists of two main components: a probabilistic surrogate model that hypothesizes about the unknown objective function, and an acquisition function that guides the selection of the most promising design candidates based on the surrogate model’s predictions and associated uncertainty. Unlike many deep learning–based design methods used in materials design \cite{Ma2019, Noh2019, Bastek2023, jin2023mechanical, Xue2020}, which typically require extensive datasets, BO operates efficiently with a small initial dataset followed by sequentially sampling and evaluating only the most promising designs. As a result, it identifies solutions with much fewer design evaluations compared to other approaches, making it both time- and resource-efficient for complex material design problems.

In many material design scenarios, designers start with only limited knowledge about where optimal solutions are located in the design space. Hence, BO with an expandable design space becomes especially valuable, as it enables exploration beyond the initially defined parameter ranges. However, such an approach has not yet been applied in materials design. To address this gap, we propose a BO design framework that incorporates a diminishing expansion rate, allowing for controlled exploration outside the initial design space. We apply this framework to a two-dimensional (2D) nacre-mimetic model system, using a five-parameter bi-linear traction-separation law to define the interface constitutive behavior. The effectiveness of the BO framework is demonstrated through two case studies. In the first scenario, a known ``ground truth” solution allows for direct validation of the method. In the second, the ground truth is unknown, yet the BO framework still successfully identifies non-unique but optimal interface designs associated with distinct deformation and failure mechanisms.

\section*{RESULTS}\label{sec:results}

\subsection*{Model Material and Interface Law} \label{sec:model}

\begin{figure}[b!]    
    \centering
    \includegraphics[width=\textwidth]{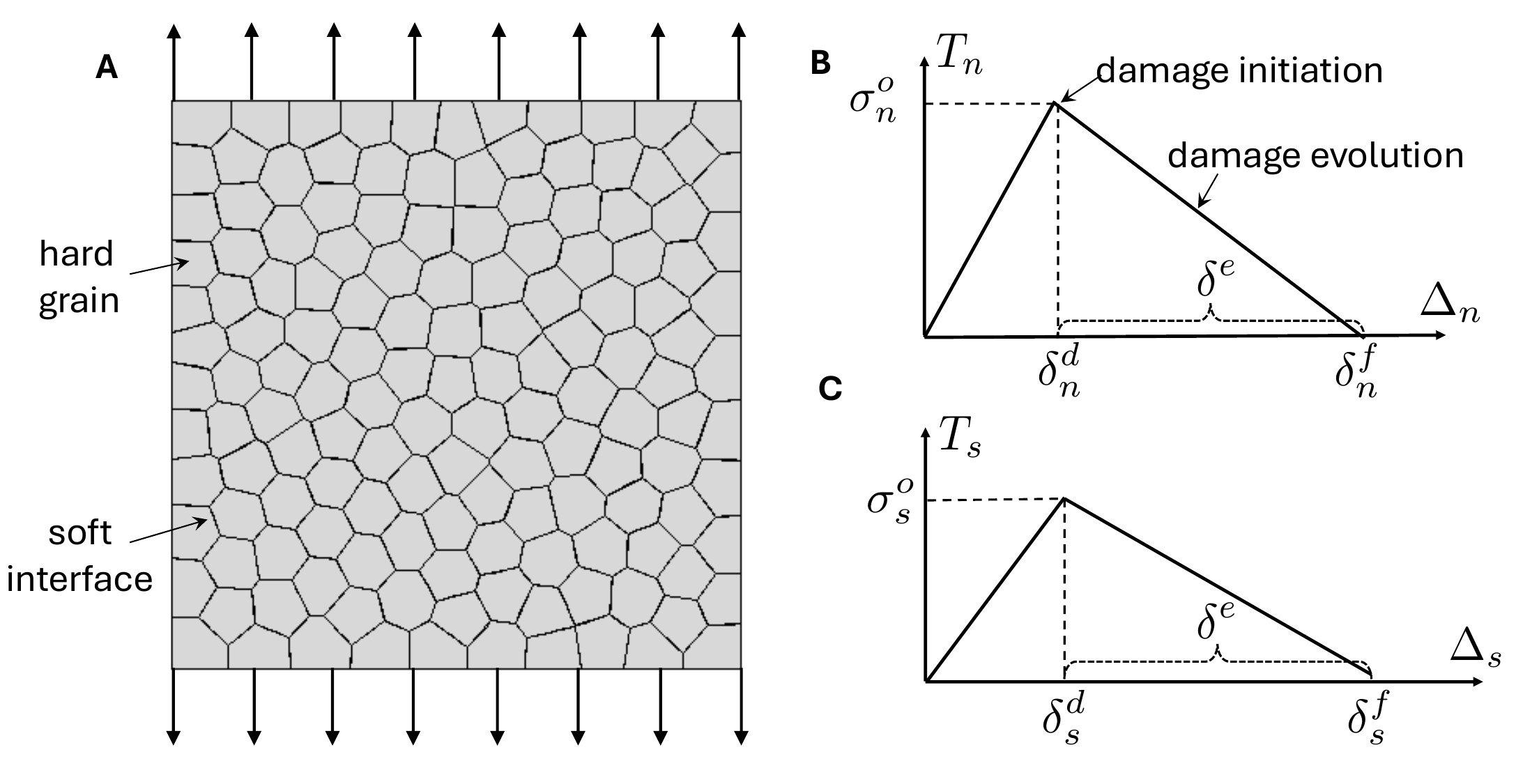}
    \caption{\textbf{The two-dimensional model material and interface law.} (A) The model is composed of hard grains glued by soft interface materials; (B) The bilinear traction-separation interface law in the normal direction; (C) The bilinear traction-separation interface law in the shear direction.
    }\label{fig:model_interface}
\end{figure}

In this study, we employ a two-dimensional model material, illustrated in Fig.~\ref{fig:model_interface}A, to demonstrate the proposed inverse design method. The structure, generated via Voronoi tessellation \cite{Rousseau2005}, consists of hard, polygonal grains bonded together by a thin adhesive layer of negligible thickness. This configuration can be viewed as a single layer of nacre and also serves as a representative model of the bone extrafibrillar matrix \cite{han2021removal}. In our analysis, we fix the material properties of the hard grains and vary the interface parameters. Under tensile loading, applied through displacement control, a corresponding tensile stress-strain curve is computed for each set of interface parameters using the finite element method (FEM).

In FEM, the interface is modeled using a bi-linear traction-separation relation \cite{Camanho2002}. As illustrated in Fig.~\ref{fig:model_interface}B and C, two bi-linear curves are used to define the interracial behavior in normal or shear directions, respectively. Five independent interface parameters are involved: $\{\sigma_{n}^{o}, \sigma_{s}^{o}, \delta_{n}^{d}, \delta_{s}^{d}, \delta^{e} \}$, where $n$ and $s$ represent the normal and shear directions. Interface strength is governed by $\sigma_{n}^{o}$ and $\sigma_{s}^{o}$, while damage initiation separations are denoted as $\delta_{n}^{d}$ and $\delta_{s}^{d}$. The parameter $\delta^{e}$ describes the progression of damage from initiation to complete failure. Damage initiation occurs when traction reaches the interfacial strength in either direction, represented by the following condition:
\begin{equation}
    max\left\{ \frac{<T_n>}{\sigma_{n}^{o}}, \frac{T_s}{\sigma_{s}^{o}} \right\}=1,
\label{eq:failure_condition}
\end{equation}
\noindent where $T_n$ and $T_s$ denote the tractions in normal and shear directions, respectively, and the symbol $< >$ stands for the Macaulay bracket, which signifies that a pure compressive deformation does not initiate damage. To describe the evolution of damage under a mixture of normal separation, $\Delta_n$, and shear separation, $\Delta_s$, an effective displacement is introduced as 
\begin{equation}
    \Delta_{m}=\sqrt{<\Delta_{n}>^2 + \Delta_{s}^2}.
\label{eq:effective_displacement}
\end{equation}
Once damage occurs, the extent of damage under mixed loading is solely determined by a damage variable $D$, which is defined as
\begin{equation}
    D=\frac{\Delta_{m}^{f}(\Delta_{m}-\Delta_{m}^{d})}{\Delta_{m}(\Delta_{m}^{f}-\Delta_{m}^{d})},
\label{eq:damage}
\end{equation}
where $\Delta_{m}^{d}$ denotes the effective displacement at which damage initiates, and $\Delta_{m}^{f}$ represents the separation at which interface fails, given by $\Delta_{m}^{f} = \Delta_{m}^{d} + \delta^{e}$. To simulate the linear softening of a cohesive element due to damage, a correction factor of $(1-D)$ is multiplied to the undamaged stress under the same separation.

Beyond the interface law, the mechanical behavior of this model composite material also depends on the density of polygonal grains. However, since the current study focuses primarily on interface design, density is not considered as a design variable. In addition, as demonstrated by the geometry sensitivity analysis (see Supporting Information), the composite’s stress-strain response remains insensitive to the variations in grain distribution at the current grain density.

\subsection*{Inverse Design Framework} \label{sec:design_framework}

\begin{figure}[t!]
    \centering
    \includegraphics[width=\textwidth]{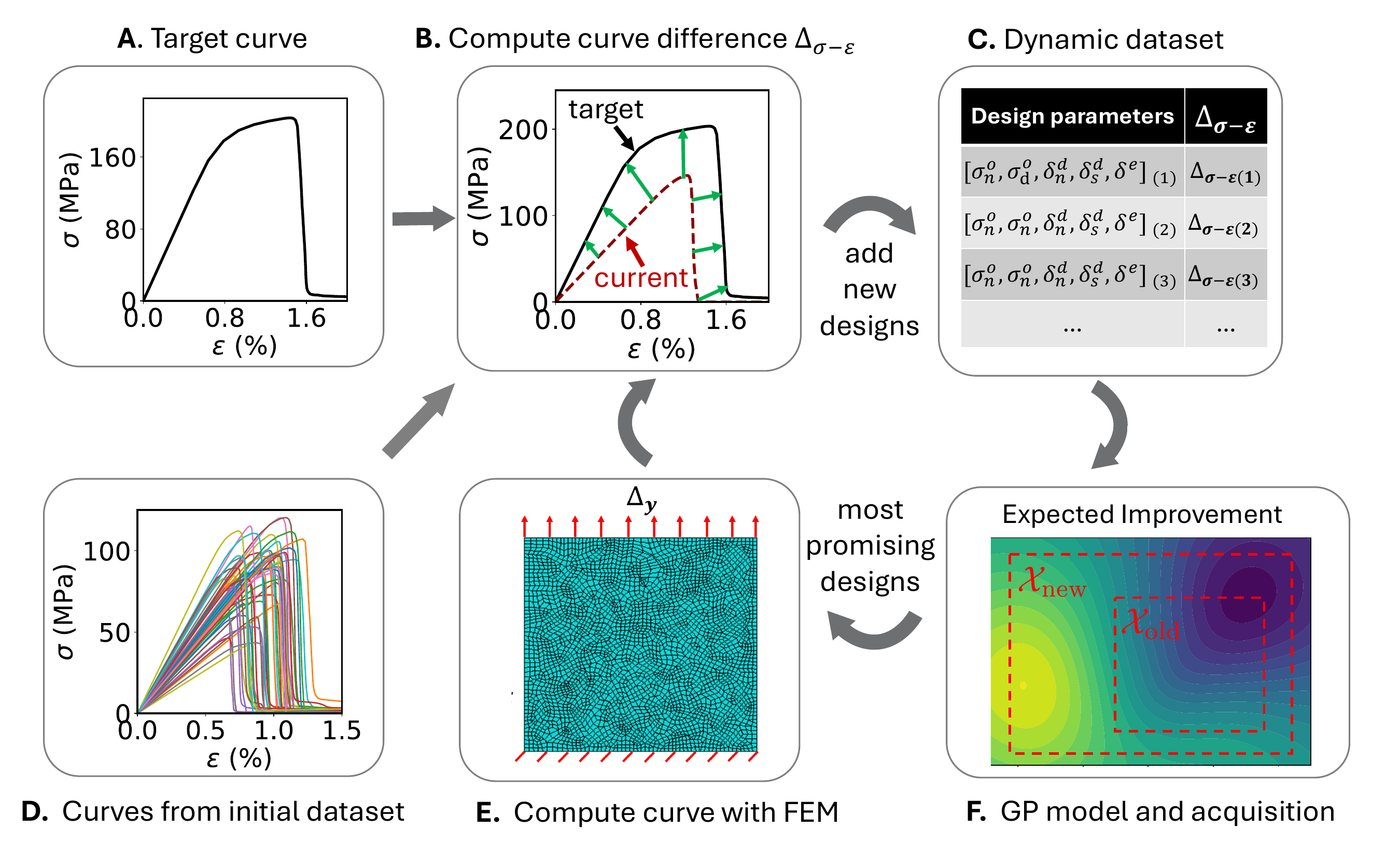}
    \caption{\textbf{Bayesian optimization design framework.} (A) A predefined target stress-strain curve; (B) The curve difference, represented by $\Delta_{\sigma-\epsilon}$, is incorporated into the objective function and is calculated based on the pointwise distances between the target curve and the curves of candidate solutions simulated by the FEM; (C) A dataset, consisting of design parameters as input features and the curve differences $\Delta_{\sigma-\epsilon}$ as labels, is iteratively augmented during the BO process; (D) The stress-strain curves from the initial dataset generated prior to the optimization process; (E) The FEM model for computing the stress-strain curve; (F) An illustration of design space expansion.}
    \label{fig:inverse_design_framework}
\end{figure}

The proposed inverse design framework is illustrated in Fig.~\ref{fig:inverse_design_framework}. Our goal is to determine one or multiple optimal sets of interface parameters that yield a desired macroscopic stress-strain curve. As shown in Fig.~\ref{fig:inverse_design_framework}A, a target stress-strain curve is provided prior to the design cycle. To start the inverse design process, an initial dataset of 50 stress-strain curves is generated using FEM, with each curve corresponding to a distinct set of interface parameters. These curves are presented in Fig.~\ref{fig:inverse_design_framework}D. A key step in the inverse design is quantifying the difference between the target curve and the FEM simulated curves, as illustrated in Fig.~\ref{fig:inverse_design_framework}B. To this end, we introduce a curve difference metric, $\Delta_{\mathcal{TS}}$, defined as 
\begin{equation}
    \Delta_{\mathcal{TS}}=\frac{1}{N_\mathcal{T}}\sum_{i=1}^{N_\mathcal{T}} D_{\mathcal{T} \rightarrow \mathcal{S}}^i + \frac{1}{N_\mathcal{S}} \sum_{i=1}^{N_\mathcal{S}} D_{\mathcal{S} \rightarrow \mathcal{T}}^i.
\label{eq:AveD}
\end{equation}
Here, $D_{\mathcal{T} \rightarrow \mathcal{S}}^i$ is the minimum distance from the $i$-th point on the target curve $\mathcal{T}$ to the simulated curve $\mathcal{S}$, and $D_{\mathcal{S} \rightarrow \mathcal{T}}^i$ is the minimum distance from the $i$-th point on the simulated curve to the target one. The parameters $N_\mathcal{T}$ and $N_\mathcal{S}$ represent the total number of points on the target and simulated curves, respectively. Both curves are discretized into equally distributed points by uniform interpolation along the curve.

This proposed metric accounts for all the differences between points on the target and simulated curves, ensuring that discrepancies across the entire curve are considered. By contrast, the Hausdorff distance \cite{Huttenlocher1993}, a commonly used image comparison metric conceptually similar to our approach, considers only the maximum difference between a point on one curve and its nearest point on the other. While the Hausdorff distance highlights the worst-case deviation, it may overlook more subtle variations between the curves, as shown in our study.

The dataset structure used in our design framework is illustrated in Fig.~\ref{fig:inverse_design_framework}C. Here, the input parameters represent the interface properties, and the label is $\Delta_{\mathcal{TS}}$. These curves are randomly generated within the predefined bounds of the initial design space. The Bayesian optimization (BO) algorithm, illustrated in Fig.~\ref{fig:inverse_design_framework}F, includes two key components: a probabilistic surrogate model and an acquisition function.  We employ Gaussian Process Regression \cite{Schulz2018} as the surrogate model to approximate the relationship between the interface parameters and $\Delta_{\mathcal{TS}}$, and we select the Expected Improvement criterion~\cite{jones1998efficient} as the acquisition function. This function outputs the expected value regarding how much $\Delta_{\mathcal{TS}}$ could be reduced from the current best result, thus guiding the search of interface parameters toward better alignment with the target stress-strain curve.

Once the most promising set of interface parameters are identified by optimizing the acquisition function, they are fed into the FEM simulation, generating a new stress-strain curve, as shown in Fig.~\ref{fig:inverse_design_framework}E. This completes one iteration of the BO loop. After each iteration, the dataset is augmented with the newly evaluated parameters and their corresponding curve difference metric. The process continues until a specified computation budget is met. Upon completing all iterations, designers can either select the best-performing design directly or evaluate multiple top candidates, taking into account additional factors such as manufacturability and cost.

Unlike conventional BO algorithms, our approach allows the design space to expand dynamically during the search for extrapolated interface parameters. This feture is critical for two reasons. First, we do not know whether the optimal solutions lie within the initially defined design space. Second, expanding the design space enhances BO’s capacity to discover more solutions, which may exhibit distinct deformation and failure mechanisms, as discussed later in the paper. To enhance computational efficiency, we employ a diminishing expansion factor, allowing broad exploration during the early iterations and more focused exploitation in subsequent stages. This dynamic expansion is illustrated in Fig.~\ref{fig:inverse_design_framework}F, where a 2D heatmap represents the expected improvement contours after extending the design space from $\mathcal{X}_{\text{old}}$ to $\mathcal{X}_{\text{new}}$. By maximizing the expected improvement based on the current Gaussian Process (GP) model, we identify multiple optimal solutions within the expanded space. A detailed description of the BO algorithm in an expandable design space is provided in the Methods section.

\begin{figure}[!htbp]
    \centering
    \includegraphics[width=0.87\textwidth]{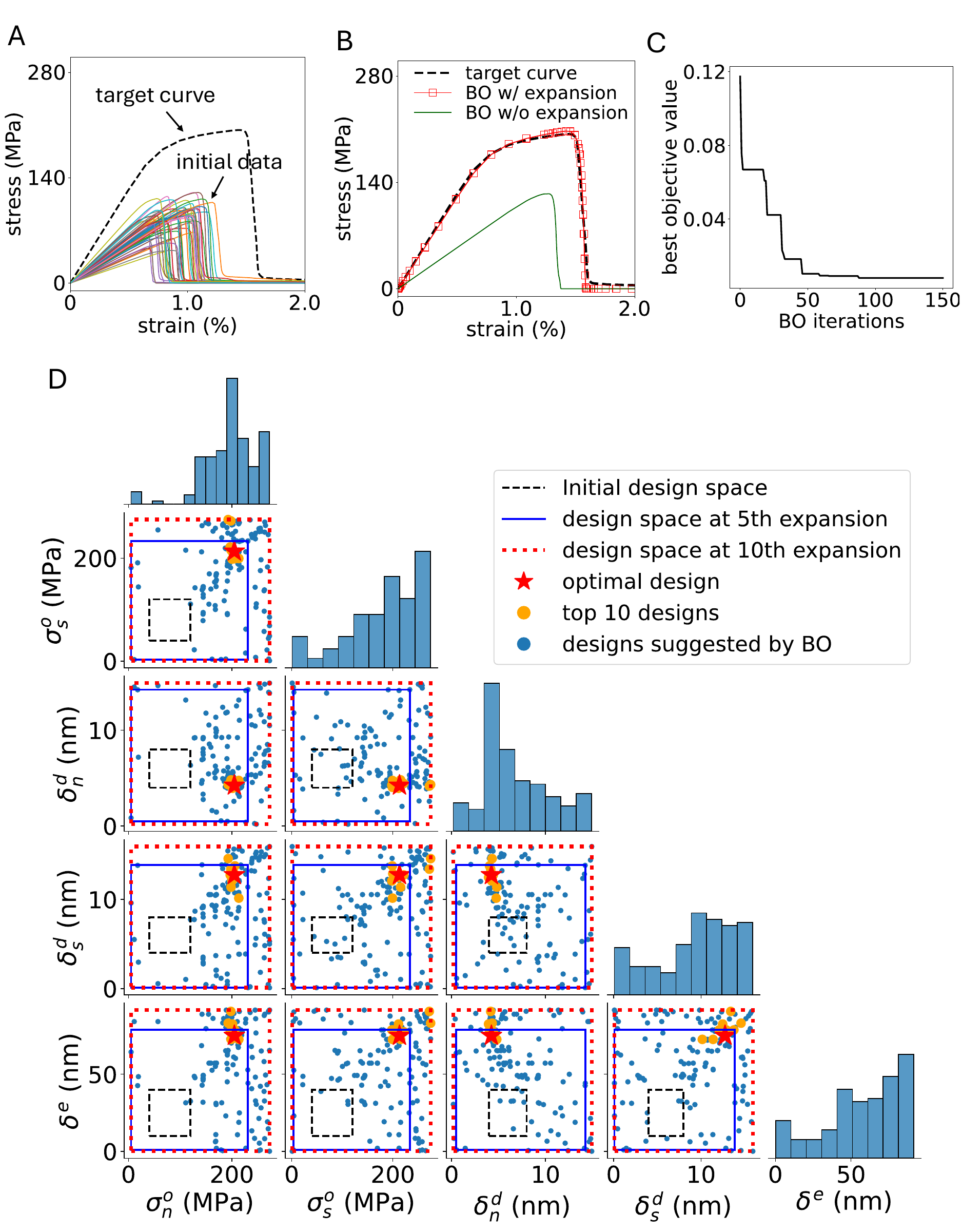}
    \caption{\textbf{Validation of BO in expandable design space.} (A) The target stress-strain curve is plotted alongside the initial curves in training data; (B) The solutions from BO with and without the design space expansion; (C) The evolution of the best objective value during BO iterations; (D) A pair plot illustrating the distribution of BO-suggested designs and the expansion of the design space, with each subplot featuring a pair of design parameters as its axes.
    }
    \label{fig:validation}
\end{figure}

\subsection*{Validation of BO in Expandable Design Space}\label{sec:validation}

To validate the proposed inverse design method, a target stress-strain curve was generated using FEM with a set of predefined interface parameters. These parameters were intentionally set outside of the initial design space to test the effectiveness of design space expansion. As a result, the target curve falls outside the range of the curves from the initial dataset, as shown in Fig.~\ref{fig:validation}A, indicating that the target curve corresponds to either an uncommon or non-existing material design inside the initial design space. The BO optimization results, both with and without the design space expansion feature enabled, are compared alongside the target curve in Fig.~\ref{fig:validation}B. The solution obtained with space expansion aligns closely with the target curve. The corresponding interface parameters are listed in Table.~\ref{tab:validation}, showing that those from the expanded design space compare closely to the parameters from target curve. 

A pair plot shown in Fig.~\ref{fig:validation}D is used to visualize the design distribution and the expansion of the design space. Each blue dot in the figure represents a design suggested by the acquisition function, associated with a specific pair of parameter values. The diagonal histograms show the distribution of designs according to each parameter. The rectangles in the figure illustrate the boundaries of the design space at its initial state, as well as after the fifth and tenth expansions. In this case, design space expansion occurs a total of ten times. The red stars indicate the location of the best solution, while the yellow dots highlight the top ten designs. The distribution of suggested design points is densely clustered around the best solution. Meanwhile, the sampling at the corners of the rectangles in a few iterations indicates a well-maintained balance between exploitation and exploration.

One key advantage of BO is its efficiency in identifying material design solutions with a relatively small number of iterations, thereby reducing the number of FEM calculations required to locate the solution for the current study. Despite setting a computational budget of 150 FEM calculations, the incremental improvements in the objective function (curve difference, $\Delta_{\mathcal{TS}}$) plotted in Fig.~\ref{fig:validation}C plateau after 45 iterations, Specifically, the improvement shifts marginally from 0.010 at the 46th iteration to 0.008 at the end. This finding suggests that computation resources could be saved by implementing a stopping criterion, such as halting the BO process once improvements remain below a predefined threshold for a certain number of iterations. Nonetheless, maintaining a larger computational budget also has its advantages. Additional iterations enable the discovery of more diverse interface designs, which may display distinct deformation and failure mechanisms, as shown in the following section.

\begin{table}[t!]
\centering 
\caption{\normalsize Interface parameters for the validation study.}
\label{tab:validation}
\begin{tabular}{lccccc} 
\toprule
& $\sigma_{n}^{o} (MPa)$ & $\sigma_{s}^{o} (MPa)$ & $\delta_{n}^{d} (nm)$ & $\delta_{s}^{d} (nm)$ & $\delta^{e} (nm)$\\ 
\midrule
Range in initial data & 40-120 & 40-120 & 4-8 & 4-8 & 10-40 \\
Target curve & 214.42 & 159.16 & 3.98 & 12.38 & 69.40 \\
BO w/o expansion & 120 & 120 & 8 & 8 & 40 \\
BO w/ expansion & 199.60 & 167.17 & 4.48 & 10.14 & 70.46 \\
\bottomrule
\end{tabular}
\end{table}

\begin{figure}[!htbp]
    \centering
    \includegraphics[width=0.8\textwidth]{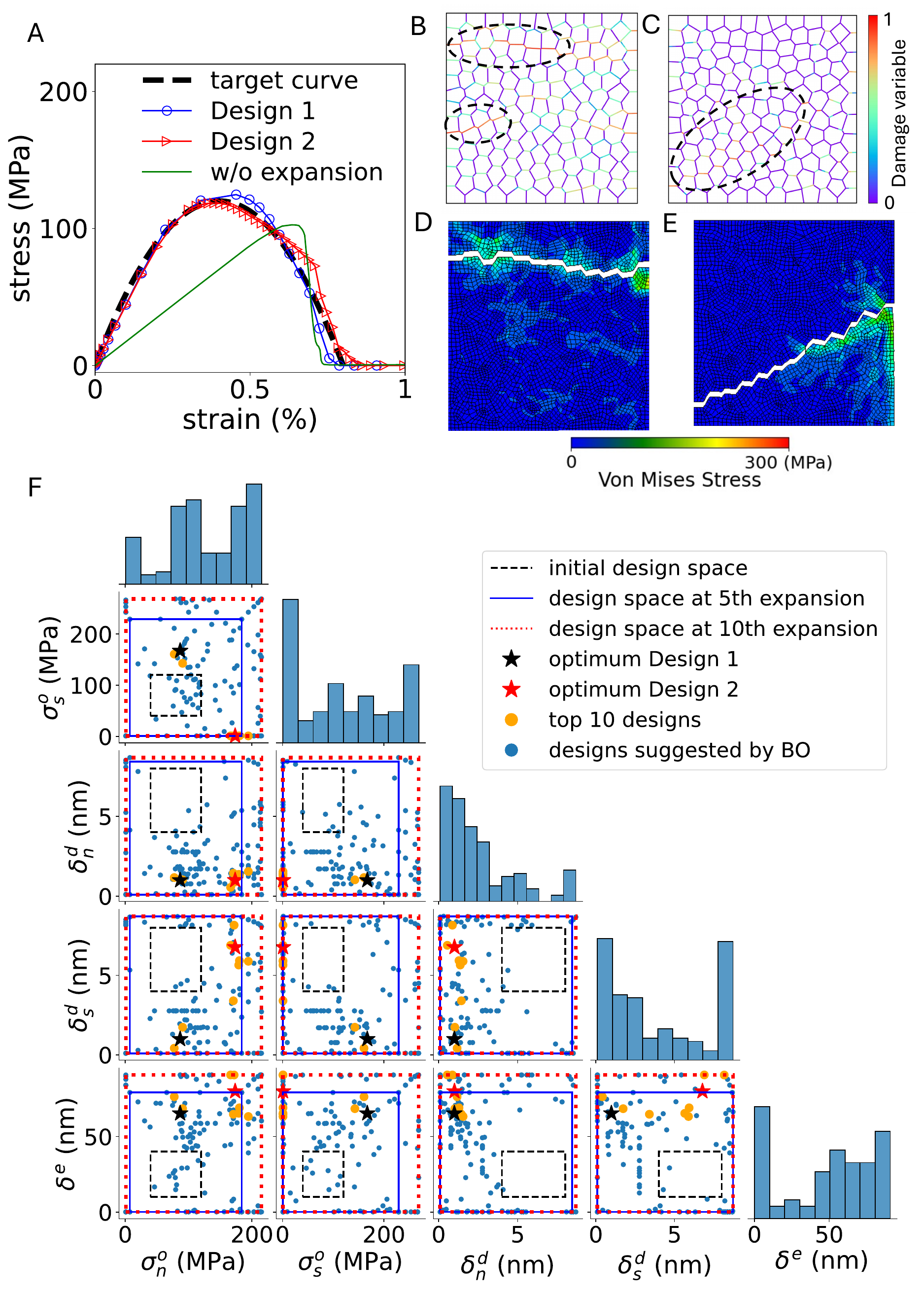}
    \caption{\textbf{Two distinct interface designs identified by BO.} (A) Two solutions representing distinct failure modes are compared to the target curve; (B, C) The interface damage distribution at the peak stress for Design 1 and 2, respectively; (D, E) The cracked FEM configuration for Design 1 and 2, respectively; (F) A pair plot illustrating the distribution of two groups of BO-suggested designs.
    }
    \label{fig:FigureFour}
\end{figure}

\subsection*{Non-unique Interface Designs for a Target Stress-strain Curve} \label{sec:nonunique}

In this section, we demonstrate that by using BO within an expandable design space, it is possible to identify multiple, non-unique interface designs that achieve a similar target stress-strain curve. Although these designs yield nearly indistinguishable mechanical responses, they differ significantly in their deformation and failure mechanisms. From a practical standpoint, having multiple design options that meet the same performance criteria provides greater flexibility during the fabrication stage and allows for more tailored solutions to meet specific application requirements.

The target curve, illustrated in Fig.~\ref{fig:FigureFour}A, is defined as a simple quadratic function. Unlike the previous validation example, this target curve is not derived from FEM, so the corresponding interface parameters are initially unknown. Notably, the shape of this target curve deviates more significantly from the curves within the initial design space compared to the earlier example, thereby potentially posing a more challenging optimization task for the BO process.

The distribution of BO-generated designs is illustrated in the pair plot shown in Fig.~\ref{fig:FigureFour}F.  Different from the previous example, the top ten solutions are more widely dispersed across the design space. Notably, these top performers naturally separate into two distinct groups based on their interface normal strength ($\sigma_{n}^{o}$) and shear strength ($\sigma_{s}^{o}$), suggesting the presence of two different deformation mechanisms. By contrast, both the critical normal separation for damage initiation ($\delta_{n}^{d}$) and the effective range of damage evolution ($\delta^{e}$) cluster within a narrow band, highlighting their pivotal influence on achieving the target response. Meanwhile, the critical shear separation for damage initiation ($\delta_{s}^{d}$) is broadly distributed, suggesting that, in this particular case, the stress-strain response is relatively insensitive to variations in $\delta_{s}^{d}$. 
From the diagonal histograms in the pair plot, it is observed that most designs are concentrated near the optimal values, as well as along the boundaries of the design space, indicating a well-managed balance between exploitation and exploration by the BO algorithm. The interface parameters and corresponding stress-strain curves for the top 10 design solutions are available in the Supporting Information.

To demonstrate BO’s capacity to identify distinct yet optimal solutions, we selected two solutions — Design 1 and Design 2 — from the top ten designs suggested by BO. As shown in Fig.~\ref{fig:FigureFour}A, the stress-strain curves from both designs closely match the target curve. However, their respective interface parameters, summarized in Table.~\ref{tab:nonunique}, are significantly different. The interface shear strength ($\sigma_s^o$) of Design 2 is two orders of magnitude lower than that of Design 1. In compensation, Design 2 accommodates more extended slip ($\delta_s^d$) between the grains before interface failure. By contrast, the interface of Design 1 exhibits a lower normal strength ($\sigma_n^o$) than Design 2. The interface parameter differences suggest that Designs 1 and 2 are prone to distinct failure mechanisms, with the former more likely to exhibit normal-dominated failure, and the latter more inclined toward shear-driven failure.

The contours of damage variable $D$ defined in Eq. \eqref{eq:damage} at the peak of the stress-strain curves of these two designs are compared in Fig.~\ref{fig:FigureFour}B and C. Design 1 exhibits regions of damage approaching the maximum value (close to 1, indicating interface failure) concentrated horizontally along the top and through the middle-left area, as highlighted in the Fig.~\ref{fig:FigureFour}B. By contrast, Design 2 exhibits a more widespread, inclined distribution of damage with generally lower damage values. 
The final fracture patterns of these two designs are shown in Fig.~\ref{fig:FigureFour}D and E. Design 1’s failure manifests as a horizontal crack, indicating normal-dominated fracture modes. Design 2, on the other hand, shows an inclined crack, consistent with shear-dominated fracture. Together, these results demonstrate how BO in an expandable design space can reveal different interface designs that achieve the same target response.

\begin{table}[t!]
\centering 
\caption{\normalsize Interface parameters of Design 1 and Design 2.}
\label{tab:nonunique}
\begin{tabular}{lccccc} 
\toprule
& $\sigma_{n}^{o} (MPa)$ & $\sigma_{s}^{o} (MPa)$ & $\delta_{n}^{d} (nm)$ & $\delta_{s}^{d} (nm)$ & $\delta^{e} (nm)$\\ 
\midrule
Design 1 & 86.71 & 167.15 & 0.99 & 0.99 & 65.38 \\
Design 2 & 173.51 & 1.00 & 1.00 & 6.78 & 80.02 \\

\bottomrule
\end{tabular}
\end{table}

\section*{DISCUSSION}
After identifying the optimal interface designs from the BO framework, the next crucial step is to realize these designs in actual materials. This requires selecting or engineering adhesive-like polymers capable of achieving the targeted interface parameters. Molecular dynamics (MD) simulation is a powerful tool for examining how polymer-surface interactions influence interface behavior at the molecular scale \cite{Zhang2016, Otsuki2021}. Moreover, machine learning approaches have shown promise in linking specific polymer structures directly to interface properties \cite{Shi2022}, enabling predictions of how variations in polymer composition affect interfacial performance. A key challenge lies in establishing a direct connection between molecular structure and the macro-scale stress-strain response. Achieving this goal will require integrating MD simulations, ML-driven structure-property models, and BO-guided macro-scale optimization into an inverse design framework.

The material model utilized in this study is predominantly governed by interfaces, with the grains themselves undergoing minimal deformation. Such problems can also be modeled using the discrete element method (DEM), which has been employed to investigate mesoscale mechanisms in nacre-like structures \cite{Abid2018} and to examine the nonlinear behavior of three-dimensional face-centered cubic (FCC) granular crystals \cite{Karuriya2024}. The BO framework proposed in this study can be seamlessly integrated with a variety of computation methods including DEM. Furthermore, the method presented here is not restricted solely to interface design; it can be extended to a wide range of material optimization tasks aimed at achieving specific nonlinear mechanical responses.

The 2D single-layer model presented in this study offers a valuable starting point for examining interfacial behavior and demonstrating the effectiveness of the proposed BO design framework. However, this simplified model has limitations compared to more complex 3D, multi-layer configurations. In particular, the 2D model cannot capture the shear interactions crucial for the sliding mechanisms between layers, nor can it reproduce the characteristic staggered layering patterns observed in natural nacre. Addressing these shortcomings will require more sophisticated 3D finite element models. Moreover, additional interface parameters can be introduced into the traction-separation law to accommodate a broader range of interface behaviors, as suggested by \cite{Lin2017}.

In this study, FEA has served as the ``ground truth” against which the BO framework can discover interface designs. Nevertheless, the validation of these computational predictions depends on experimental verification. As additive manufacturing techniques continue to advance, fabricating the proposed 2D models is becoming increasingly feasible \cite{ko2020impact}. A promising direction for future work is to integrate computational optimization with experimental validation and scalable manufacturing methods, thereby bridging the gap between theoretical design and practical implementation.

The BO algorithm proposed in this study can identify diverse interface designs that produce similar target stress-strain responses by expanding the design space. However, there is no guarantee that a wide range of interface designs can be obtained within a fixed computational budget. Although, in principle, it is likely to yield distinct designs by investing computational resources and emphasizing exploration in the BO process, but this approach becomes prohibitively expensive as the training data increases. For many practical applications, achieving an almost match to the target stress-strain curve (like the examples in this study) is not strictly necessary. Instead, having a substantial set of designs that approximate the target behavior reasonably well can provide valuable flexibility and adaptability in interface fabrication. Moving forward, it would be highly beneficial to develop inverse design methods, coupled with rigorous uncertainty estimation, that can rapidly generate diverse ``good enough'' interface designs.

\section*{METHODS}
\subsection*{Algorithm of Bayesian Optimization in Expandable Design Space}
In this study, we developed a Bayesian optimization (BO) algorithm capable of expanding the design space in a controlled manner, enabling the discovery of suitable interface parameters for achieving a specified target stress-strain curve. The algorithm, detailed in Algorithm~\ref{alg: 1}, combines the advantages from previous studies \cite{Nguyen2017, pmlr-v51-shahriari16}. The degree of design space expansion is governed by a scaling parameter $\gamma$. During each expansion, the span of design space, $\boldsymbol{h}_t$, is scaled by a factor of $\sqrt[d]{\frac{\gamma}{N}}$, where $N$ denotes the current count of expansions and $d$ is the dimension of the design space. In this work, we adopt a dynamic scaling parameter $\gamma=N+1$, resulting in pronounced expansions in early iterations that gradually taper off as the search progresses. This down-scaling strategy enhances computational efficiency by promoting broad exploration in early iterations and targeted exploitation in later stages.

\begin{figure}[!htbp]
    \centering
    \includegraphics[width=0.5\textwidth]{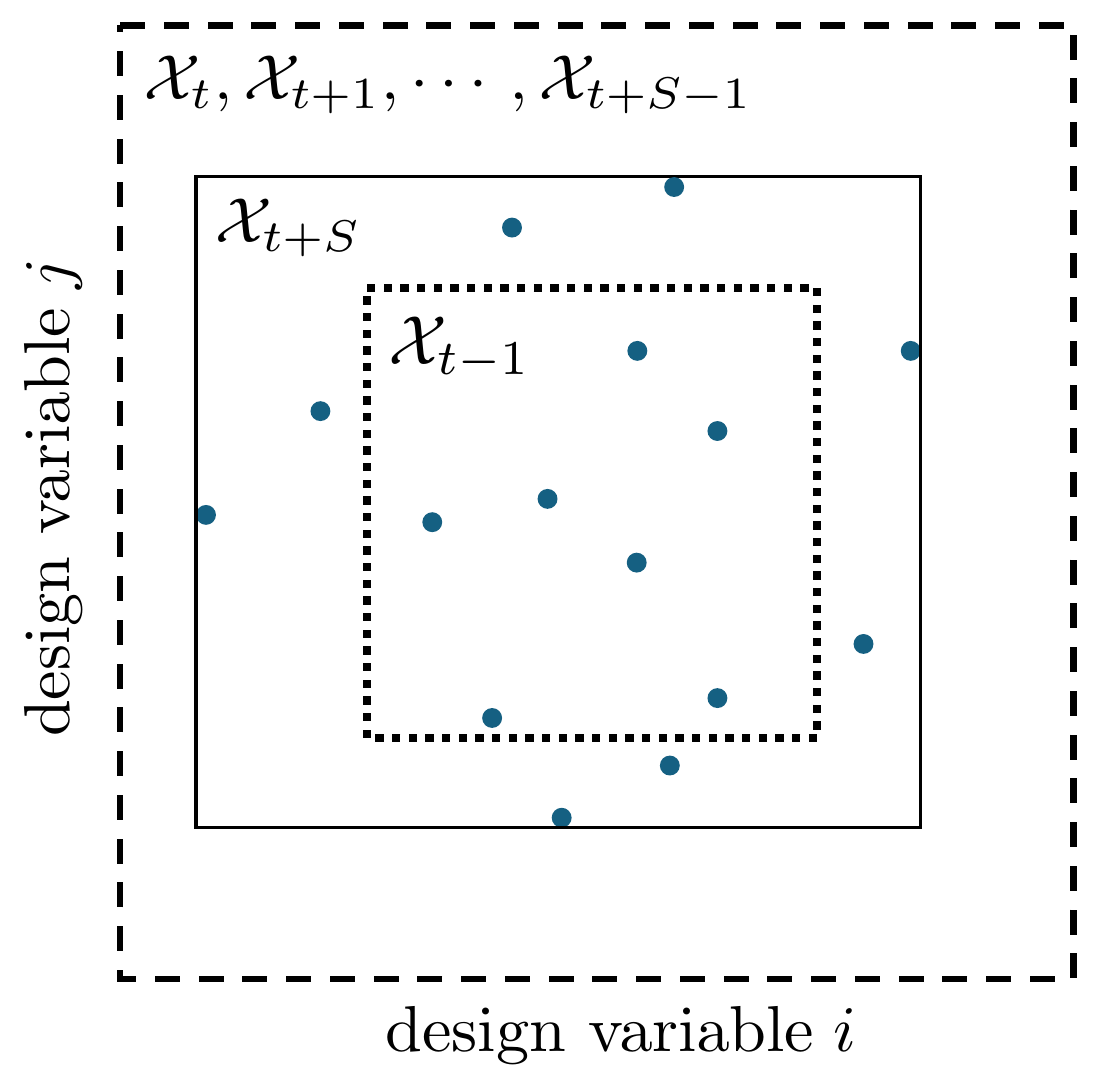}
\caption{\textbf{Illustration of design space expansion.}}
    \label{fig:space_expansion}
\end{figure}

\begin{algorithm}[!ht]
\caption{Bayesian optimization in expandable design space}
\label{alg: 1}
\begin{algorithmic}
\State \textbf{Input:} maximum iterations $T$; stall number $S$; initial dataset $\boldsymbol{D}_{0}$; initial design space $\mathcal{X}_0$ with spans $\boldsymbol{h}_{0}$; dimension of design space $d=5$; scale parameter $\gamma$.
\begin{algorithmic}[1] 
\State Set expansion count $N=0$
\For{$t = 1$ to $T$}
    \State Fit a Gaussian Process model $\mathcal{GP}_t$ based on the dataset $\boldsymbol{D}_{t-1}$
    \If{$t\%S==1$, }
        \State $N=N+1$
        \State Expand the design space to $\mathcal{X}_{t}$ with spans $\boldsymbol{h}_{t}=\boldsymbol{h}_{t-1}\times \sqrt[d]{\frac{\gamma}{N}}$
        \State Acquire $\boldsymbol{x}_t=\underset{x\in \mathcal{X}_{t}}{\text{argmax }} EI_{t}(x;\mathcal{GP}_{t})$, where $EI$ is the expected improvement 
    \Else{, \textbf{do}}
        \State $\mathcal{X}_{t}=\mathcal{X}_{t-1}$
        \State Acquire $\boldsymbol{x}_t=\underset{x\in \mathcal{X}_{t-1}}{\text{argmax }} EI_{t}(x;\mathcal{GP}_{t})$
        \If{$t\%S==0$, }
                \State Update $\mathcal{X}_t$ based on $\mathcal{X}_{t-S}$ and the recently acquired $S$ designs $\{\boldsymbol{x}_{t-S+1}, \ldots, \boldsymbol{x}_{t}\}$ 
        \EndIf
    \EndIf
    \State Run FEM, compute curve difference, $\Delta_{\mathcal{TS}}^t$ using Eq.~\eqref{eq:AveD}
    \State Augment the dataset $\boldsymbol{D}_t=\boldsymbol{D}_{t-1}\cup{(\boldsymbol{x}_t, \Delta_{\mathcal{TS}}^t)}$
\EndFor
\end{algorithmic}
\State\textbf{Output:} All solutions $\{\boldsymbol{x}_1, \boldsymbol{x}_2,\cdots,\boldsymbol{x}_T\}$
\end{algorithmic}
\end{algorithm}

The expansion of design space is illustrated in Fig.~\ref{fig:space_expansion}. The dotted-line rectangle represents the boundary of the design space at BO iteration $t-1$, denoted by $\mathcal{X}_{t-1}$. After expansion, the design space extends to the dashed-line rectangle according to the prescribed  scaling law. Following this expansion, $S$ consecutive BO iterations, referred to as the ``stall” phase, are conducted within this expanded space. During this phase, the design spaces ($\mathcal{X}_t, \mathcal{X}_{t+1}, \ldots, \mathcal{X}_{t+S-1}$) remain fixed, and BO selects $S$ new designs (shown as solid dots in the figure) by maximizing the expected improvement. After the stall ends, the design space is updated to the minimal region that encompasses both the previous space $\mathcal{X}_{t-1}$ and the $S$ newly selected designs, resulting in a reduced and more focused area denoted by $\mathcal{X}_{t+S}$, illustrated as the solid-line rectangle in the figure.This space refinement is aimed at directing computational effort toward the most promising areas, thereby enabling more efficient and accurate identification of high-performing interface designs. The expansion and stall phases alternate until all iterations are completed. At the end, designers can directly select the best solution identified by BO or examine several top candidates, considering additional factors such as manufacturability and cost-effectiveness.  

\subsection*{Finite Element Model}
2D geometric models are generated using Voronoi tessellation \cite{Rousseau2005} to mimic the structure of a single-layer of nacre composite. Then the geometric model is imported to ABAQUS to compute the macroscopic stress-strain curves. In the FEM model, the CPS4R elements, which are 4-node bi-linear plane stress quadrilaterals, are used for the grains, and the COH2D4 cohesive elements are adopted for the interface, which exhibit zero geometric thickness but non-zero constitutive thickness in calculation. In FEM model, the lower edge of the structure is fixed in all the translational and rotational directions, and a prescribed displacement is assigned to the upper edge step by step until fracture occurs. For each interface design, the output of FEA is a stress-strain curve starting from the initially undeformed configuration to the final prescribed displacement.


\section*{Data and Code Availability}
{\raggedright
The datasets used to train the model, as well as the source code, are available at 
\href{https://github.com/Gao-Group/BO-interface-inverse-design}{https://github.com/Gao-Group/BO-interface-inverse-design}.
\par}

\section*{Acknowledgments}
W.G. and W.C. gratefully acknowledge the Texas A\&M Engineering Experiment Station startup funds. In addition, W.G. acknowledges the Governor’s University Research Initiative. This research were conducted with the advanced computing resources provided by Texas A\&M High Performance Research Computing.

\section*{Author Contributions}
\textbf{Wei Zhang}: Methodology, Software, Investigation, Writing - original draft, Writing - review \& editing. \textbf{Mingjian Tang, Haoxuan Mu, Xiaowei Zeng, Rui Tuo, Wei (Wayne) Chen}: Investigation, Writing - review \& editing. \textbf{Wei Gao}: Conceptualization, Methodology, Software, Investigation, Writing - original draft, Writing - review \& editing, Supervision, Funding acquisition.

\section*{Conflict of Interest}
The authors declare no conflict of interest.

\clearpage 

%
\bibliography{ref} 
\bibliographystyle{sciencemag}

%
%
%
%
%
%



\newpage


\renewcommand{\thefigure}{S\arabic{figure}}
\renewcommand{\thetable}{S\arabic{table}}
\renewcommand{\theequation}{S\arabic{equation}}
\renewcommand{\thepage}{S\arabic{page}}
\setcounter{figure}{0}
\setcounter{table}{0}
\setcounter{equation}{0}
\setcounter{page}{1} 


\begin{center}
\section*{Supplementary Materials for\\ \scititle}

    Wei Zhang,
    Mingjian Tang,
    Haoxuan Mu,\\
    Xingzi Yang,
    Xiaowei Zeng$^{\ast}$,
    Rui Tuo,\\
    Wei (Wayne) Chen$^{\ast}$,
    Wei Gao$^{\ast}$\\
\small$^\ast$Corresponding author. wei.gao@tamu.edu (Wei Gao), xiaowei.zeng@utsa.edu (Xiaowei Zeng), w.chen@tamu.edu (Wei (Wayne) Chen)
\end{center}

\subsubsection*{This PDF file includes:}
Supplementary Note 1-2

\noindent Table S1 and S2

\noindent Figure S1 to S3

\newpage
\subsection*{Grain geometry sensitivity study}

Two-dimensional geometric models are generated using Voronoi tessellation to mimic the structure of a single-layer nacre-like composite, where the polygonal grains represent the stiff components, and the grain boundaries serve as biopolymeric interfaces. Under this approach, the grain density (i.e., the total number of grains per unit area) remains constant, while their positions and shapes vary, yielding different grain patterns. Although low grain densities can make the macroscopic response more sensitive to grain arrangements, the chosen grain density in this study produces minimal variation.

\begin{figure}[!htbp]
    \centering
    \includegraphics[width=0.9\linewidth]{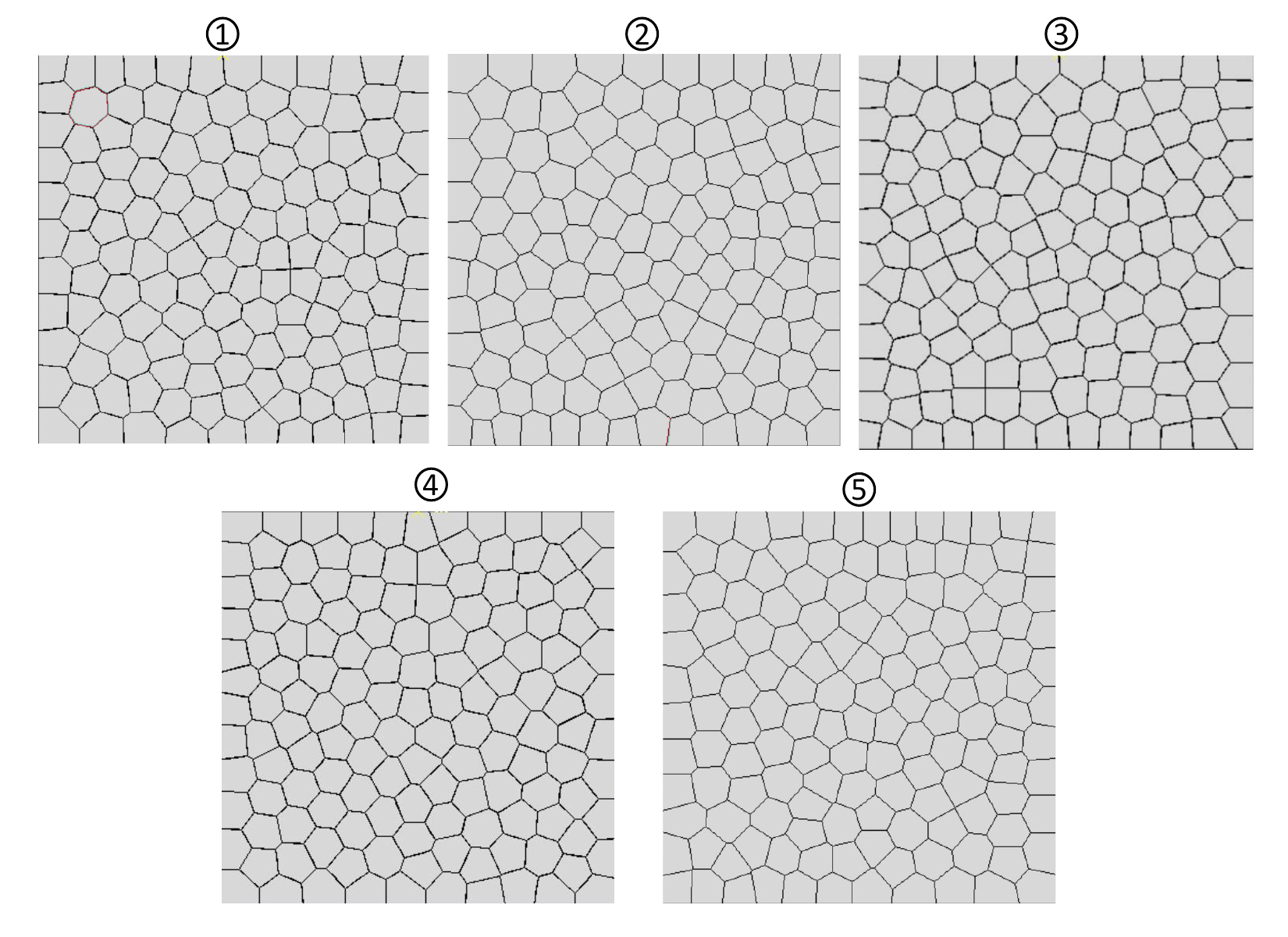}
    \caption{A set of grain patterns, all with the same grain density, used to investigate the sensitivity of the stress-strain curve to geometric variability.}
    \label{fig:grain_geometry}
\end{figure}

To assess the impact of grain pattern variation, we evaluated five distinct patterns with three sets of interface material properties listed in Table~\ref{tab:mat_geometry_sensitivity}. As illustrated in Fig.~\ref{fig:grain_geometry}, the minimal variation among the resulting stress-strain curves indicates that the composite’s macroscopic response is relatively insensitive to the specific grain arrangement at the current grain density. Consequently, geometry 4 was selected for the subsequent study.

\begin{table}[!b]
\centering 
\caption{\normalfont Parameter values of the three materials used in sensitivity study.}
\label{tab:mat_geometry_sensitivity}
\begin{tabular}{lccccc} 
\toprule
& $\sigma_{n}^{o} (MPa)$ & $\sigma_{s}^{o} (MPa)$ & $\delta_{n}^{d} (nm)$ & $\delta_{s}^{d} (nm)$ & $\delta^{e} (nm)$\\ 
\midrule
material 1 & 125 & 125 & 2 & 2 & 15 \\
material 2 & 200 & 125 & 4 & 2 & 15 \\
material 3 & 125 & 200 & 2 & 4 & 15 \\

\bottomrule
\end{tabular}
\end{table}

\begin{figure}[!htbp]
    \centering
    \includegraphics[width=1\linewidth]{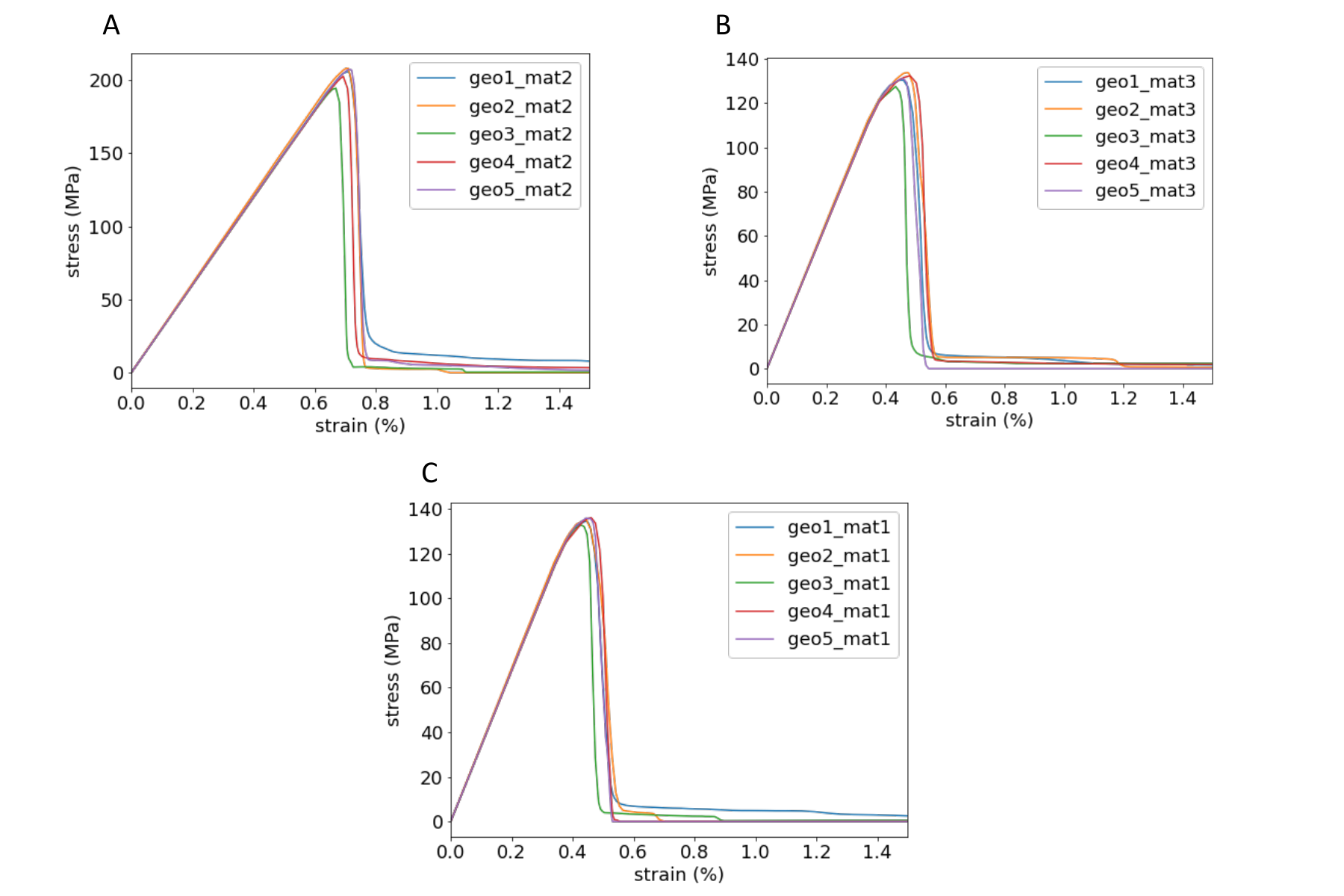}
    \caption{Comparisons of the stress-strain curves for three different interface materials. In each case, five distinct grain patterns are examined, while the interface material remains unchanged.}
    \label{fig:curves_test_geometry}
\end{figure}

\newpage

\subsection*{Supplementary Note 2. Non-unique interface designs}
\begin{figure}[!ht]
    \centering
    \includegraphics[width=0.8\linewidth]{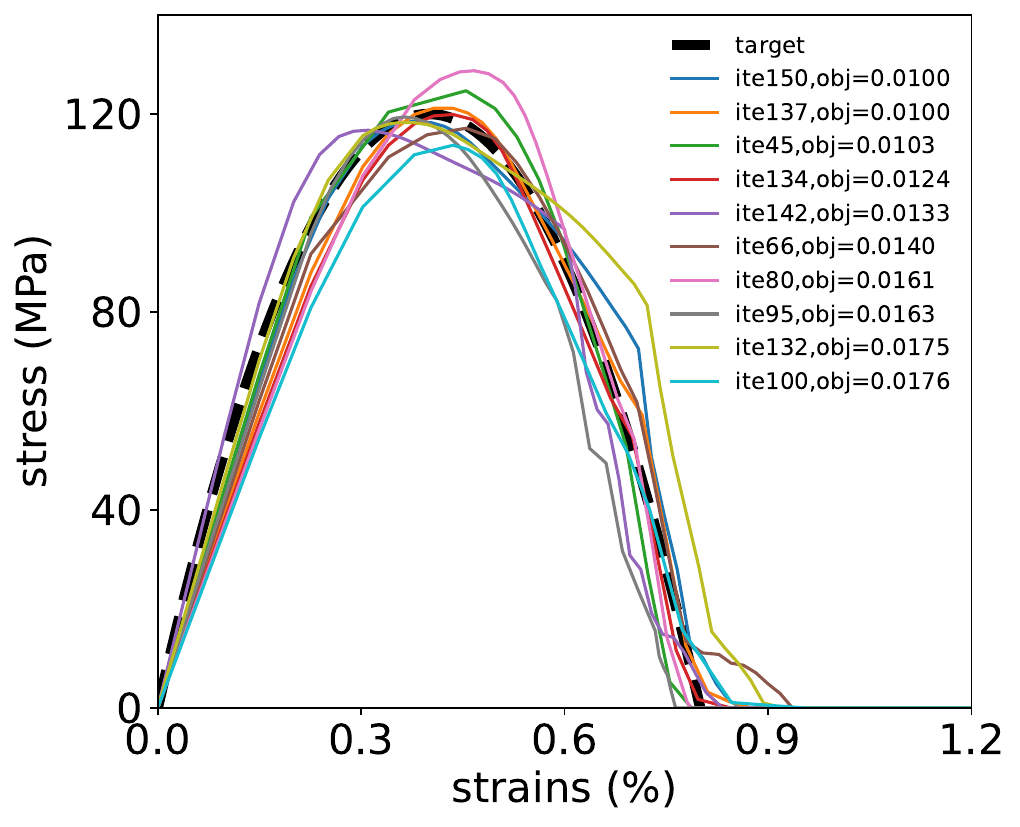}
    \caption{The stress-strain curves of the top ten interface designs in the second inverse design example discussed in the main text.}
    \label{fig:10_best}
\end{figure}

In the second inverse design example discussed in the main text, we demonstrated BO’s capability to identify multiple distinct interface designs that yield similar stress-strain responses. Table~\ref{tab:interface_10_best} lists the top ten designs suggested by BO, and their corresponding stress-strain curves are presented in Fig.~\ref{fig:10_best}. These ten designs were selected at different iterations based on their objective (curve difference) values, as shown in the figure.

Notably, the interface parameters split into two distinct groups: one group features high normal interface strength and fails predominantly by shear, while the other group exhibits high shear strength and fails under normal loading. This result highlights BO’s ability not only to find optimal solutions that approximate the target curve but also to uncover fundamentally different failure mechanisms.

\begin{table}[!t]
\centering 
\caption{\normalfont Interface parameters top ten interface designs in the second inverse design example discussed in the main text.}
\label{tab:interface_10_best}
\begin{tabular}{lccccc} 
\toprule
& $\sigma_{n}^{o} (MPa)$ & $\sigma_{s}^{o} (MPa)$ & $\delta_{n}^{d} (nm)$ & $\delta_{s}^{d} (nm)$ & $\delta^{e} (nm)$\\ 
\midrule
1 & 173.51 & 1.00 & 1.00 & 6.78 & 80.02 \\
2 & 180.26 & 1.00 & 1.29 & 5.93 & 69.08 \\
3 & 86.71 & 167.15 & 0.99 & 0.99 & 65.38 \\
4 & 179.12 & 1.00 & 1.37 & 5.65 & 65.43 \\
5 & 166.32 & 1.00 & 0.55 & 6.89 & 90.87 \\
6 & 90.61 & 142.74 & 1.02 & 1.74 & 68.65 \\
7 & 194.32 & 1.00 & 1.55 & 5.89 & 63.32 \\
8 & 77.43 & 160.75 & 1.17 & 0.42 & 76.74 \\
9 & 171.88 & 1.00 & 0.87 & 8.16 & 90.82 \\
10 & 170.67 & 1.00 & 1.44 & 3.40 & 64.82 \\

\bottomrule
\end{tabular}
\end{table}

\end{document}